\documentclass[final,authoryear,5p,twocolumn]{elsarticle}

\usepackage{amssymb}

\begin{document}


\def\aj{\rm{AJ}}                   
\def\aap{\rm{A\&A}}                
\def\mnras{\rm{MNRAS}}             

\journal{Astronomy and Computing}


\begin{frontmatter}



\title{PARAVT: Parallel Voronoi Tessellation code}


\author[myad1,myad2]{Roberto E. Gonz\'alez}

\address[myad1]{Instituto de Astrof\'{i}sica, Pontificia Universidad Cat\'olica,
  Av. Vicu\~na Mackenna 4860, Santiago, Chile}
\address[myad2]{Centro de Astro-Ingenier\'{i}a, Pontificia Universidad Cat\'olica,
  Av. Vicu\~na Mackenna 4860, Santiago, Chile}
\ead{regonzar@astro.puc.cl}

\begin{abstract}
We present a new open source code for massive parallel computation of Voronoi tessellations(VT hereafter) in large data sets.
The code is focused for astrophysical purposes where VT densities and neighbors are widely used.
There are several serial Voronoi tessellation codes, however no open source and parallel implementations are available to handle the large number of particles/galaxies in current N-body simulations and sky surveys.

Parallelization is implemented under MPI and VT using Qhull library. Domain decomposition takes into account consistent boundary computation between tasks, and includes periodic conditions.

In addition, the code computes neighbors list, Voronoi density, Voronoi cell volume, density gradient for each particle, and densities on a regular grid.

Code implementation and user guide are publicly available at 

{\it https://github.com/regonzar/paravt}.
\end{abstract}


\end{frontmatter}


\section{Introduction}
\label{sintro}

The Voronoi Tessellation \citep{voronoi} technique define a cellular-like structure, where each particle is associated to a region (or Voronoi cell) in which any point inside this region is nearest to that particle than to any other. 

In the field of Astrophysics, in particular for N-body simulations, this is a very useful tool to identify immediate neighbors of particles and it is one of the best adaptive methods to recover a precise density field from a discrete distribution of points, with clear advantage over Smoothed Particle Hydrodynamic or other interpolation based techniques \citep{2000A&A...363L..29S, 2003A&A...403..389P},  
where its principal asset is its complete independence of arbitrary smoothing functions and parameters specifying the properties of these. VT reproduce the anisotropies of the local particle distribution and through its adaptive and local nature proves to be optimally suited for uncovering the full structural richness in the density distribution.
Other remarkable uses of VT in this field are filamentary structure identification \citep{2010MNRAS.407.1449G}, N-body simulation code AREPO \citep{2010MNRAS.401..791S}, { halo and void identification \citep{2005MNRAS.356.1222N,2008MNRAS.386.2101N}, and non-parametric determination of halo concentrations \citep{2015ApJ...811..152L}.}

There are several serial open source VT implementations such as Qhull\footnote{{\it http://www.qhull.org}}
 \citep{Barber:1996:QAC:235815.235821}, or Voro++\footnote{{\it http://math.lbl.gov/voro++}} \citep{voro}.
{ ZOBOV/VOBOZ algorithms use VT for halo and void finding in cosmological data sets \citep{2005MNRAS.356.1222N,2008MNRAS.386.2101N}, they divide data in sub-volumes to process them serially, however this code is not implemented in parallel, and does not output all VT structure data.}
{ CGAL library\footnote{\it http://www.cgal.org} also computes VT and can run in parallel using threads, however this implementation does not take into account buffer zones for proper VT computation at sub-volumes boundaries, and it works only on shared memory architectures (single nodes), where memory becomes a bottleneck and scaling is limited when compared with distributed architectures under MPI.}
{ \citet{9780471986355} show an extensive review on VT computation methods.}
There are also parallel algorithms for VT computation such as \citet{2014CoPhC.185.3204S}, however none of them are open source and freely available to the community.
{ Another parallel VT implementation is tess2\footnote{{\it https://github.com/diatomic/tess2}} \citep{Peterka:2014:HCD:2683593.2683702}  }, which does not take into account buffer zones and is poorly documented.

\begin{figure*}[!thb]
\begin{center}
\includegraphics[width=.95\linewidth,natwidth=1413,natheight=948,angle=0]{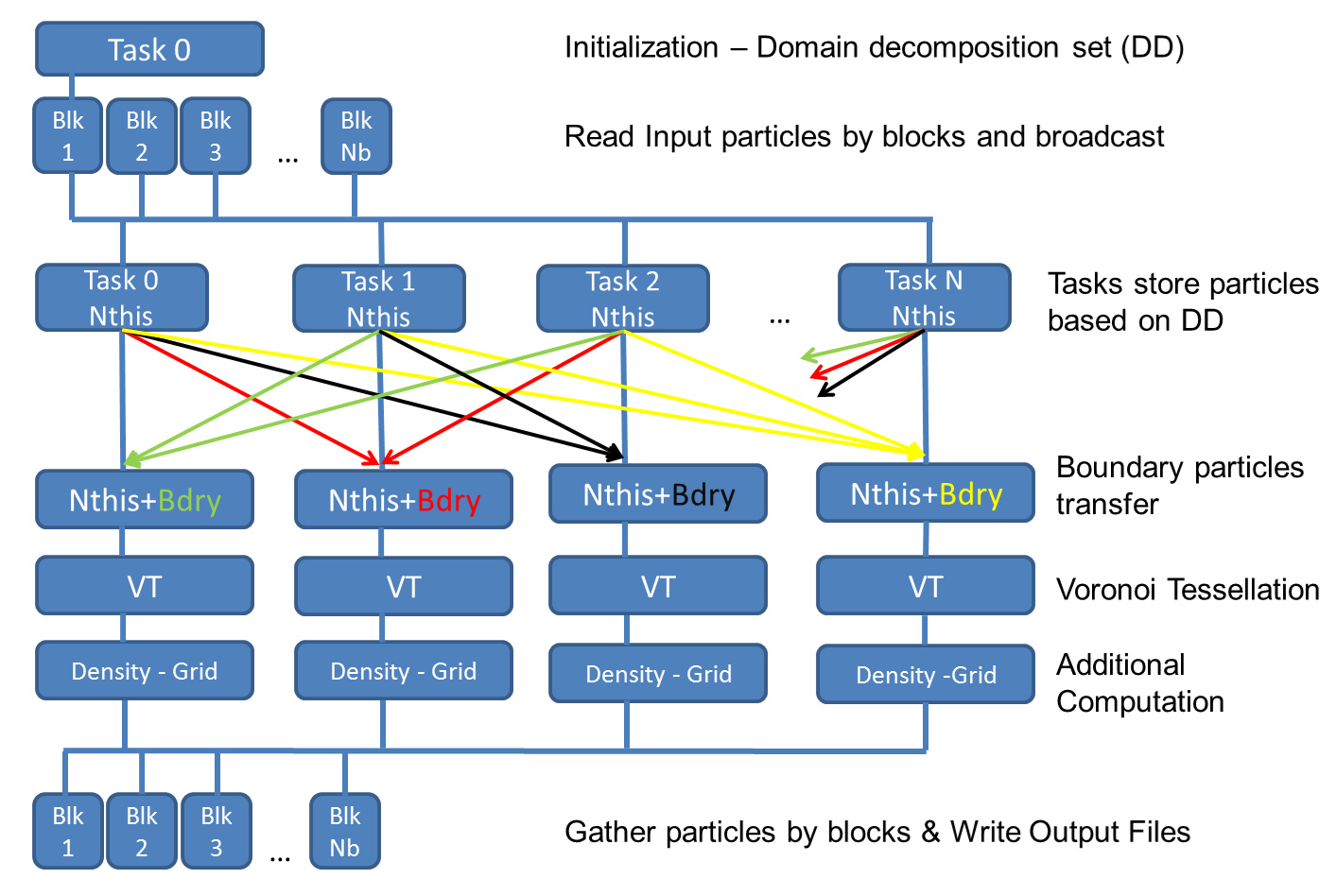}
\caption{
\label{figdiagram} Flow diagram.}
\end{center}
\end{figure*}

The purpose of this code is to be an open source tool for VT computation of large data sets, parallel and optimized to take full advantage of actual multicore and { distributed memory} cluster architectures, user-friendly with documentation, and with typical I/O formats used in the field of N-body simulations.
In addition, this code computes useful properties such as Voronoi densities, cell volumes, { density gradients}, and immediate neighbors lists.

\section{Implementation}
\label{simp}
Code is developed in C, MPI and use the re-entrant Qhull C library. It is organized in different separated modules.
A brief description of the code is shown in the flow diagram from figure \ref{figdiagram}, where code run sequentially from top to bottom, and parallelization is represented as horizontal branches.


After initialization, domain decomposition is defined depending on the number of tasks and configuration. 
There are two domain decomposition schemes, a) split the volume in { powers of $2$, balancing the cuts in all  dimensions, until there are a number of regions equal to the number of MPI tasks $N_{TASK}$, all with equal volumes}. Therefore, for $N_{TASK}$, volume is split in each dimension by $nx$, $ny$, $nz$ such as $nx\times ny\times nz=N_{TASK}$. i.e. $N_{TASK}=32$ split volume in each dimension by $nx=4$, $ny=4$, and $nz=2$. b) split the volume along a single direction $N_{TASK}$ times.
Scheme a) is most suitable for simulation boxes, and b) { is intended for irregular and elongated volumes such as galaxy survey slices and cones where the volume have a length along a particular dimension much larger than the other two\footnote{Examples of these conical elongated volumes are: The 2dF galaxy redshift survey($http://www.2dfgrs.com$), and GAMA survey ($http://www.gama-survey.org$) }.}

The input file is read by a single root task. Particles are read in blocks of a given buffer size, then each block is broadcast to the other tasks. Each task selects particles from the buffer depending on domain decomposition and allocates for them.

{ After all particles are distributed among tasks,}
each task defines which particles should be sent to neighboring tasks as boundary particles.
The thickness of the layer of boundary particles 
{ is defined by a BORDERFACTOR code parameter, it is in units of the
inter-particle distance which is computed using the whole input data.
Notice that it is not possible to know in advance the required thickness without an expensive computation, or a prior VT computation, then we need an approximation. 
Safe values for cosmological simulations range between $1$ and $3$, lower values may lead to insufficient particles for correct VT computation, and larger values may lead to an overhead of boundary particles and decrease in performance.}

\begin{figure}[!htb]
\begin{center}
\includegraphics[width=.98\linewidth,angle=0,natwidth=822,natheight=1056]{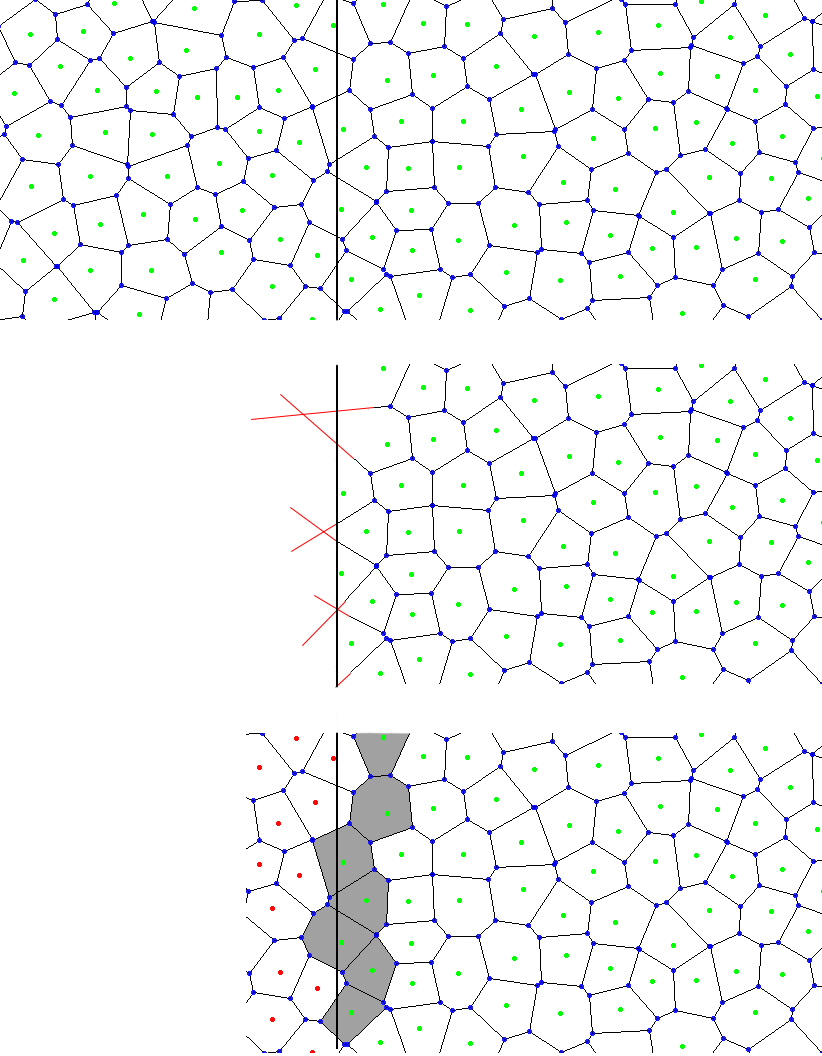}
\caption{
\label{figbound} Consistent treatment of task boundaries. Top panel) an example particle distribution in 2D with their corresponding VT. We want to split volume in two processes defined by the vertical black line. Middle panel) If we compute VT of particles at the right side, particles at the border of division will have an incorrect VT computation, see red lines. Bottom panel) if we add boundary particles(Red dots), the VT of previously mentioned particles(gray dashed) will have a correct VT.}
\end{center}
\end{figure}

In figure \ref{figbound} we show a graphical example of why boundary particles are needed when we split volume in several sub-volumes according to the domain decomposition. { The top panel shows} a particle distribution in 2D which will be split by the vertical black line. The middle panel shows how the VT of particles at the border of a division will fail if no boundary consideration is taken(red lines). In the bottom panel we add some boundary particles(red dots), such that the failed particles from middle panel (Grey dashed cells) now have a correct VT.

{
In a most general picture, there may be cases of extremely uneven particle distributions where boundary layer must be increased by a larger factor, and even reach a thickness where parallelization does not give any benefit, in such cases a different domain decomposition scheme is required.
An optimal domain decomposition must balance the load for each task and at the same time reduce the boundary surface between tasks, for instance a domain decomposition which split the volume in equal number of particles may lead to very large surfaces between tasks increasing the number of buffer boundary particles and communication. 
There are several domain decomposition schemes based on tree space partitioning such as kd-tree \citep{9783540779735}, and based on different space filling curves such as Peano-Hilbert curves \citep{9783642310461}, most of these methods produce a very well balanced space partition, but none of them guarantee a minimal boundary surface. Few methods consider a minimization of the boundary. For cosmological simulations \citet{Wu2011} propose a domain decomposition for optimal load balance and a boundary surface with minimal density to reduce communication between tasks in clustered particle distributions resulting from N-body cosmological simulations.
}

{ In this code we implemented two basic domain decomposition schemes based on split the volume in equal parts, however in some cases of extremely uneven particle distributions, these schemes may be inefficient. 
In future versions, we plan to include other domain decomposition schemes, however the code can identify if the current domain decomposition is not appropriate and we include additional features to handle most cases of uneven particle distributions.
Code warns the user against too thin boundary layer which may lead to inaccurate VT at borders, and also warns if too thick boundary layer which mean the parallelization granularity is reached or the BORDERFACTOR parameter is too large.
An alternate feature included in the code is AUTOBORDER which automatically increases the boundary layer thickness between two tasks if it is too thin for correct VT computation. 
This adaptive solution is useful for uneven particle distributions and only increases the boundary layer thickness in the required interfaces between two adjacent tasks; the drawback is a small decrease in performance.
}

\begin{figure*}[!htb]
\begin{center}
\includegraphics[width=.9\linewidth,angle=0,natwidth=900,natheight=900]{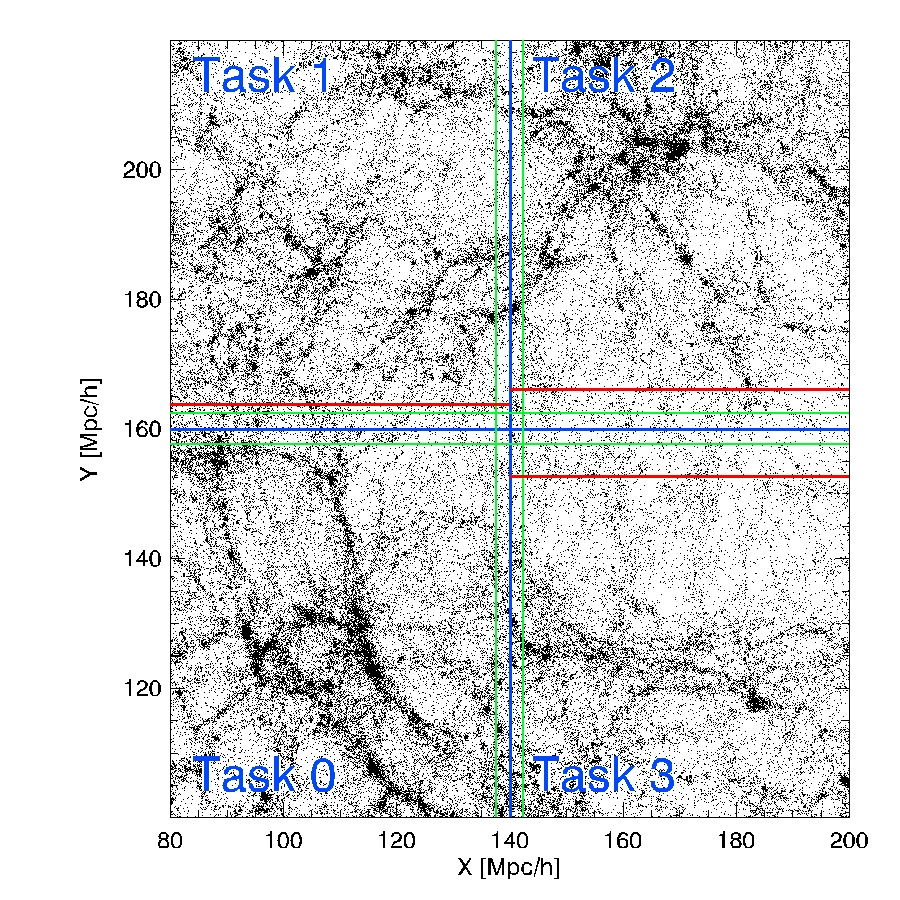}
\caption{
\label{figauto} 
{ Example of AUTOBORDER feature. 
This is a distribution of halos in a $\Lambda$CDM cosmological simulations at $z=0$.
Domain decomposition split the volume in $4$ equal regions assigned to tasks $0$ to $3$ (blue lines). 
Boundary buffer thickness defined by the inter-particle distance multiplied by a BORDERFACTOR of 1.5 is shown in green lines.
Particle density is computed in each buffer zone between tasks, and if the density is lower than the average particle density, the thickness is increased until reach the expected number of particles for the initial buffer zone volume (red lines). 
For task $2$ in the interface with task $3$ the thickness is amplified a factor of $3.1$, for task $3$ and task $0$ the amplification factors are $2.6$ and $1.6$ respectively. The other interfaces do not require amplification since they are above average density.
 }
}
\end{center}
\end{figure*}

{ Figure \ref{figauto} shows an example on how AUTOBORDER feature works in an uneven particle distribution. There is a 2D projection of the halo distribution in a $\Lambda$CDM cosmological simulations at $z=0$ made using Gadget2 code \citep{2005MNRAS.364.1105S}. 
The volume is split in $4$ equal regions using the implemented domain decomposition (blue lines). 
Tasks are labeled in each corresponding sub-volume from task $0$ to $3$. 
The load balance results in $33\%$, $23\%$, $25\%$, and $19\%$ fraction of particles allocated in tasks $0$ to $3$ respectively.
In this example we use a BORDERFACTOR of $1.5$, and the code computes an inter-particle of $1.58 h^{-1} \rm{Mpc}$, then the boundary buffer thickness is computed as $2.37 h^{-1} \rm{Mpc}$ (green lines). 
The code computes the particle density in each buffer zone between tasks, and if the density is lower than the average particle density, the thickness is increased until reach the expected number of particles for the initial buffer zone volume (Red Lines).
In this example, the buffer thickness for task $2$ in the interface with task $3$ is amplified by a factor of $3.1$, for task $3$ in the interface with task $2$ it is amplified a factor of $2.6$, and finally for task $0$ in the interface with task $1$ the amplification factor is $1.5$. The other interfaces do not require amplification since they are larger than average density.
In this particular case, the large amplification of boundary buffer zones in tasks $2$ and $3$ do not degrade overall performance, since task $0$ still contains more particles.
}

The next step, after all boundary particles are transferred, is VT computation using the Qhull library in each task. Then for each particle the code looks for all their facets to compute the Voronoi cell volume. Therefore, density is computed using particle mass and Voronoi cell volume.
{ Density gradient vector at the position of each particle is also computed using density and neighboring information.}
In addition, the code 
{ extracts the Voronoi neighbors for each particle, this can be done using Voronoi facets structure where we have that  between two neighbor particles always exist a common facet \citep{9780471986355}.}
Optionally, the code can compute average Voronoi densities on a regular grid in parallel, where each task resolves its corresponding region of the total grid.

\begin{figure*}[!htb]
\begin{center}
\includegraphics[width=.71\linewidth,angle=0,natwidth=1024,natheight=667]{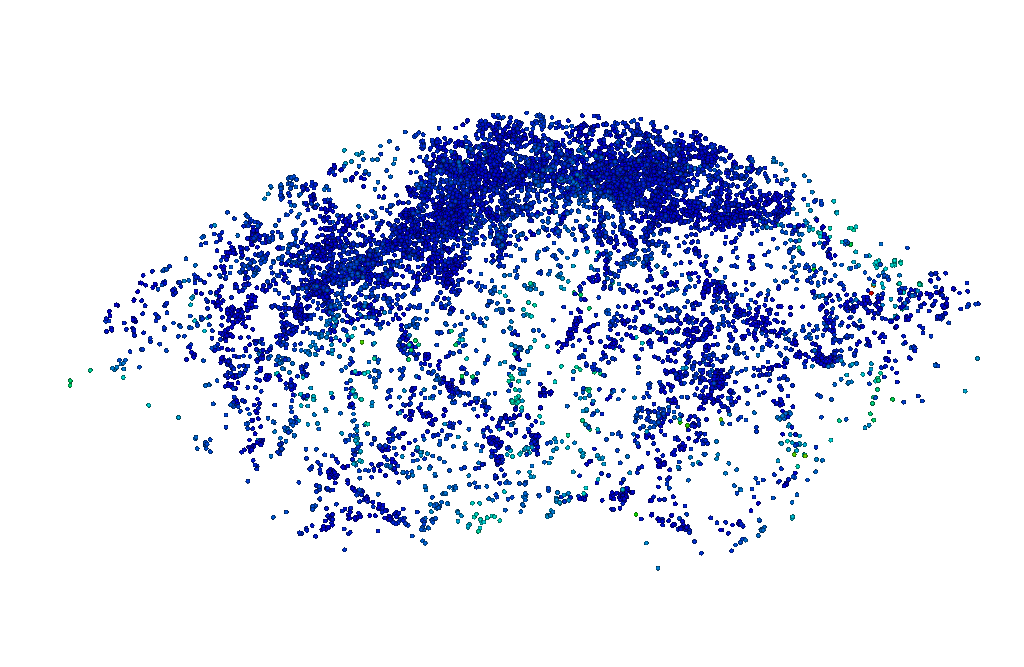}
\includegraphics[width=.71\linewidth,angle=0,natwidth=1024,natheight=667]{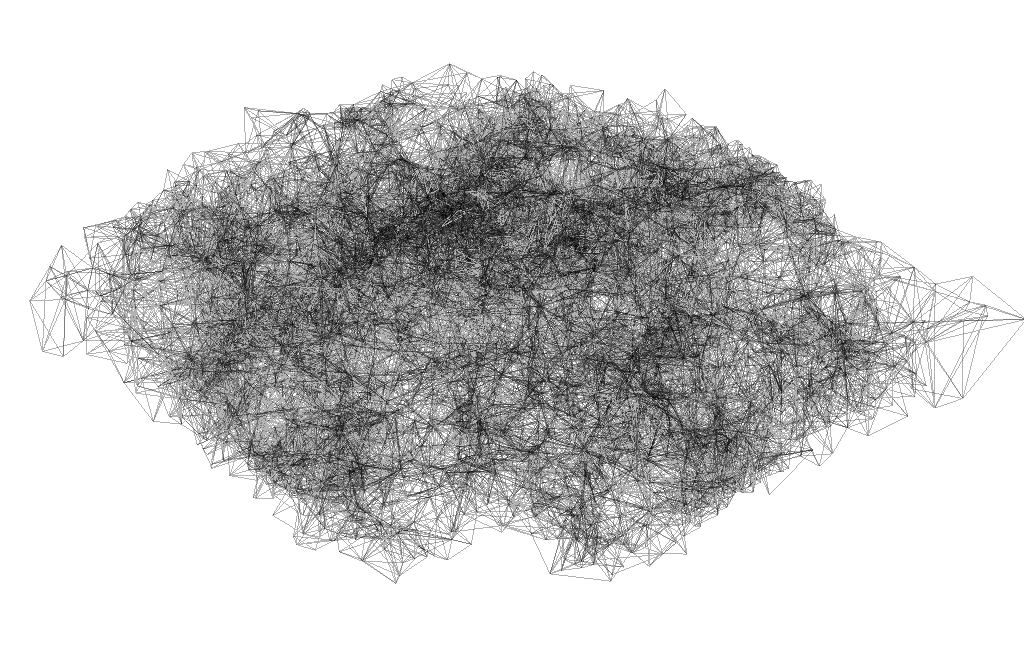}
\caption{
\label{figsample}
{ Top panel: Mock galaxy distribution for a fraction of the SDSS galaxy survey. This slice contains $\sim11000$ galaxies, colors represent galaxy mass. This is a side view of a 3D conic section. Bottom panel: VT structure of galaxy distribution, voronoi cell shapes forms a complex network useful for advanced methods in large scale structure identification. }
}
\end{center}
\end{figure*}

After all tasks finish computation, 
{ the root task request output information (density, volume, neighbors) from the other tasks.
This request is made following the same file order from input data and in blocks of particles defined by a communication buffer.
Each task checks if it has the particles requested by the root task, if yes, its output information is transmitted back to the root task.
The root task gather all output information from current requested block, and write to disk. VT neighbor information is re-indexed to replace boundary particles indices by real indices.
The gathering and writing process finish after all block are written.
 }

{ This merging process end up with output data having the same order as from input data, and neighbors data with indices following same order.}

In figure \ref{figsample} we show the VT structure output of a small galaxy distribution produced by this code, { where the number of galaxies is $\sim 11000$ and resemble a sub-volume of the SDSS galaxy survey\footnote{{\it http://www.sdss.org/}} \citep{2000AJ....120.1579Y}}.


\section{Performance}
\label{sperf}

The nice part of VT parallelization is that computation is very local. For a given particle, its VT depends only on its neighbors, and no long range computations are required { such as for gravity computation, however in some cases with extremely uneven particle distributions, neighbors may be located far from particles. In general, excluding extreme cases,} when we split the volume in different tasks, even if we add an overhead of boundary particles, total computation time is strongly reduced. 

Qhull runs VT in $O(N_P \log{N_V})$ for a typical 3D particle distribution, where $N_P$ is the number of particles, and $N_V$ is the number of vertices. 
{ The number of vertices depends on particle distribution. In typical $\Lambda$CDM cosmological simulations following \citet{2014A&A...571A..16P} cosmological parameters, with boxes having sizes $40$, $100$, and $250 h^{-1} \rm{Mpc}$ at $z=0$ we found the average number of vertices per particle is $\sim 7-10$. In other distributions such as a random Poisson point process, the average number of vertices per cell is $\sim15-20$ \citep{PhysRevE.88.063309}.}
{ Therefore, if computation times goes linear with the number of particles then scaling should be close to linear as well.}

\begin{figure}[!htb]
\begin{center}
\includegraphics[width=.98\linewidth,angle=0]{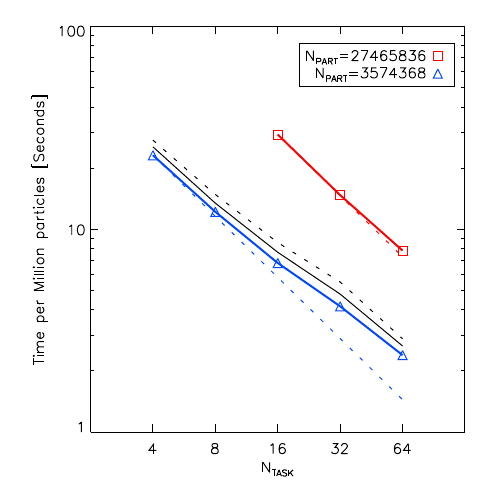}
\caption{
\label{figbench} Time benchmarks per million particles as function of the number of { MPI tasks ($N_{TASK}$)} for: a small test of $\sim3.5$ Million particles(blue), and a medium test of $\sim27.5$ Million particles(red). Dashed lines just represent a linear scaling progression. Computation times exclude I/O operations. { Black dashed lines show benchmark for the small test using $2$ and $3$ times the initial boundary thickness.}}
\end{center}
\end{figure}

We run performance tests in two nodes with following specifications:
\begin{verbatim}
SuperServer 2U 8028B-C0R3FT
Dual Socket R (LGA 2011)
4 INTEL HASWELL-EX 14C E7-4850V3 2.2G 35M 8GT/s
32 16GB DDR3-1600 1.35V 2Rx4 ECC REG DIMM
Mellanox ConnectX-3 VPI,
QSFP QDR IB (40Gb/s) and 10GbE
\end{verbatim} 

In figure \ref{figbench} we show scaling of the code for two cosmological N-body data sets, where the boundary thickness is defined as $1.5$ times the inter-particle distance (BORDERFACTOR parameter). { The first data set contains $\sim 3.5$M particles, and the second data set contains $\sim 27.5$M particles. $N_{PART}$ is the number of particles, and $N_{TASK}$ is the number of MPI tasks}. 
Computation times are normalized per million particles, { however the second data set takes more computation per million particles than the former data set, because the latter processes $\sim 8$ times less particles than the second data set per MPI task}. 
{ The scaling is almost linear in both test cases}. Differences from the linear case appear at larger number of tasks since there is additional communication and synchronization between tasks and the number of boundary particles increases as well. 
{ We also include results for the small test using different boundary thickness of $3.0$ (solid black) and $4.5$ (dotted black) times the inter-particle distance. In this test case when we increase the boundary thickness by a factor of $2$ results in a $\sim 15\%$ longer computation time, and for a factor of $3$ the computation time increases a $\sim 30\%$}.

Parallelization speedup degrades dramatically when the number of boundary particles becomes similar to the number of input particles.
A safe limit for parallelization is when boundary particles are lower than $50\%$ input particles, this can be regulated by the number of tasks and setting the boundary layer thickness.

\subsection{Memory Consumption}

{ The code is designed in MPI for distributed memory architectures. This is the optimal platform for massive parallelization in large computer clusters, where the main bottleneck in the VT computation of large data sets is the memory consumption.
The code consumes around $\sim3$Gb of RAM per million particles, and in the case of the full code output including full VT structure and grid computation the code may consume an extra $500$Mb of RAM per million particles. 
The load balance between tasks is given by the number of particles assigned to each task which depends on particle distribution.
The root task does not consume extra memory compared to the other tasks because it deliver and gather data based on a communication buffer and does not need to store extra particles information from other tasks.}

{
The code includes a parameter LOWMEMORY to reduce memory consumption by $\sim40\%$ but taking double computation time.
This feature just run half of the tasks first and the remaining tasks after the first half finish and release vertices and facets data.
This option must be run in a round-robin configuration or in a single node to guarantee memory reduction in all nodes uniformly.
Another configuration to reduce memory consumption is set the precision data type for Qhull computation. Default data type is $8$ bytes double, but this can be changed to $4$ bytes floats to cut memory consumption in most architectures, this does not improve computation speed.
 }

\section{Discussion}

PARAVT is a new parallel and open source code for VT computation designed to handle large number of particles. It is focused on astrophysics purposes with I/O formats compatible with common N-body processing tools. Additional properties are computed for each particle, such as Voronoi density, density gradient, cell volume and neighbors list. 

{ This MPI parallel code makes possible VT computation of large data-sets taking advantage of modern multicore and cluster architectures with distributed memory, where domain decomposition split data into several sub-volumes assigned to each MPI tasks including boundary buffer particles for proper VT computation of particles located at sub-volumes boundaries. 
After parallel VT computation, results are re-indexed following same input data order. Codes computes VT structure, cell volume, density and density gradient vector for each particle. Optionally code can compute on the fly and in parallel the average density on a regular grid. 
Input/Output data allows periodic conditions and multimass particles in ASCII or Binary format, in addition, common N-body cosmological formats are included such as Gadget2 \citep{2005MNRAS.364.1105S}, Rockstar \citep{2013ApJ...762..109B} , and VTK format\footnote{{\it http://www.vtk.org/}} for visualization. Code includes memory and precision management and some additional optimization features.
}

{ Several new features will  be included in next versions of the code, such a new domain decomposition schemes, grid output for tidal tensor and hessian matrix analysis, additional I/O formats, consistency checks for boundary zones, and more.}

{ Code is publicly available at Github repository{\footnote{{\it https://github.com/regonzar/paravt}
}}, and it includes a complete user guide. }

\section*{Acknowledgement}
REG was supported by Proyecto Financiamiento Basal PFB-06 'Centro de
Astronomia y Tecnologias Afines' and Proyecto Comit\'e Mixto ESO
3312-013-82.
The Geryon cluster at the Centro de Astro-Ingenieria UC was
extensively used for the calculations performed in this paper. The
Anillo ACT-86, FONDEQUIP AIC-57, and QUIMAL 130008 provided funding
for several improvements to the Geryon cluster.

\bibliographystyle{model2-names}
\bibliography{parabib}

\begin{thebibliography}{21}
\expandafter\ifx\csname natexlab\endcsname\relax\def\natexlab#1{#1}\fi
\providecommand{\url}[1]{\texttt{#1}}
\providecommand{\href}[2]{#2}
\providecommand{\path}[1]{#1}
\providecommand{\DOIprefix}{doi:}
\providecommand{\ArXivprefix}{arXiv:}
\providecommand{\URLprefix}{URL: }
\providecommand{\Pubmedprefix}{pmid:}
\providecommand{\doi}[1]{\href{http://dx.doi.org/#1}{\path{#1}}}
\providecommand{\Pubmed}[1]{\href{pmid:#1}{\path{#1}}}
\providecommand{\bibinfo}[2]{#2}
\ifx\xfnm\relax \def\xfnm[#1]{\unskip,\space#1}\fi
\bibitem[{Bader(2012)}]{9783642310461}
\bibinfo{author}{Bader, M.}, \bibinfo{year}{2012}.
\newblock \bibinfo{title}{Space-Filling Curves: An Introduction with
  Applications in Scientific Computing: 9 (Texts in Computational Science and
  Engineering)}.
\newblock \bibinfo{publisher}{Springer}.
\bibitem[{Barber et~al.(1996)Barber, Dobkin and
  Huhdanpaa}]{Barber:1996:QAC:235815.235821}
\bibinfo{author}{Barber, C.B.}, \bibinfo{author}{Dobkin, D.P.},
  \bibinfo{author}{Huhdanpaa, H.}, \bibinfo{year}{1996}.
\newblock \bibinfo{title}{The quickhull algorithm for convex hulls}.
\newblock \bibinfo{journal}{ACM Trans. Math. Softw.} \bibinfo{volume}{22},
  \bibinfo{pages}{469--483}.
\newblock \URLprefix \url{http://doi.acm.org/10.1145/235815.235821},
  \DOIprefix\doi{10.1145/235815.235821}.
\bibitem[{{Behroozi} et~al.(2013){Behroozi}, {Wechsler} and
  {Wu}}]{2013ApJ...762..109B}
\bibinfo{author}{{Behroozi}, P.S.}, \bibinfo{author}{{Wechsler}, R.H.},
  \bibinfo{author}{{Wu}, H.Y.}, \bibinfo{year}{2013}.
\newblock \bibinfo{title}{{The ROCKSTAR Phase-space Temporal Halo Finder and
  the Velocity Offsets of Cluster Cores}}.
\newblock \bibinfo{journal}{ApJ} \bibinfo{volume}{762}, \bibinfo{pages}{109}.
\newblock \DOIprefix\doi{10.1088/0004-637X/762/2/109},
  \href{http://arxiv.org/abs/1110.4372}{\tt arXiv:1110.4372}.
\bibitem[{de~Berg et~al.(2008)de~Berg, Cheong, van Kreveld and
  Overmars}]{9783540779735}
\bibinfo{author}{de~Berg, M.}, \bibinfo{author}{Cheong, O.},
  \bibinfo{author}{van Kreveld, M.}, \bibinfo{author}{Overmars, M.},
  \bibinfo{year}{2008}.
\newblock \bibinfo{title}{Computational Geometry: Algorithms and Applications}.
\newblock \bibinfo{publisher}{Springer}.
\bibitem[{{Gonz{\'a}lez} and {Padilla}(2010)}]{2010MNRAS.407.1449G}
\bibinfo{author}{{Gonz{\'a}lez}, R.E.}, \bibinfo{author}{{Padilla}, N.D.},
  \bibinfo{year}{2010}.
\newblock \bibinfo{title}{{Automated detection of filaments in the large-scale
  structure of the Universe}}.
\newblock \bibinfo{journal}{\mnras} \bibinfo{volume}{407},
  \bibinfo{pages}{1449--1463}.
\newblock \DOIprefix\doi{10.1111/j.1365-2966.2010.17015.x},
  \href{http://arxiv.org/abs/0912.0006}{\tt arXiv:0912.0006}.
\bibitem[{{Lang} et~al.(2015){Lang}, {Holley-Bockelmann} and
  {Sinha}}]{2015ApJ...811..152L}
\bibinfo{author}{{Lang}, M.}, \bibinfo{author}{{Holley-Bockelmann}, K.},
  \bibinfo{author}{{Sinha}, M.}, \bibinfo{year}{2015}.
\newblock \bibinfo{title}{{Voronoi Tessellation and Non-parametric Halo
  Concentration}}.
\newblock \bibinfo{journal}{ApJ} \bibinfo{volume}{811}, \bibinfo{pages}{152}.
\newblock \DOIprefix\doi{10.1088/0004-637X/811/2/152},
  \href{http://arxiv.org/abs/1504.04307}{\tt arXiv:1504.04307}.
\bibitem[{Lazar et~al.(2013)Lazar, Mason, MacPherson and
  Srolovitz}]{PhysRevE.88.063309}
\bibinfo{author}{Lazar, E.A.}, \bibinfo{author}{Mason, J.K.},
  \bibinfo{author}{MacPherson, R.D.}, \bibinfo{author}{Srolovitz, D.J.},
  \bibinfo{year}{2013}.
\newblock \bibinfo{title}{Statistical topology of three-dimensional
  poisson-voronoi cells and cell boundary networks}.
\newblock \bibinfo{journal}{Phys. Rev. E} \bibinfo{volume}{88},
  \bibinfo{pages}{063309}.
\newblock \DOIprefix\doi{10.1103/PhysRevE.88.063309}.
\bibitem[{{Neyrinck}(2008)}]{2008MNRAS.386.2101N}
\bibinfo{author}{{Neyrinck}, M.C.}, \bibinfo{year}{2008}.
\newblock \bibinfo{title}{{ZOBOV: a parameter-free void-finding algorithm}}.
\newblock \bibinfo{journal}{\mnras} \bibinfo{volume}{386},
  \bibinfo{pages}{2101--2109}.
\newblock \DOIprefix\doi{10.1111/j.1365-2966.2008.13180.x},
  \href{http://arxiv.org/abs/0712.3049}{\tt arXiv:0712.3049}.
\bibitem[{{Neyrinck} et~al.(2005){Neyrinck}, {Gnedin} and
  {Hamilton}}]{2005MNRAS.356.1222N}
\bibinfo{author}{{Neyrinck}, M.C.}, \bibinfo{author}{{Gnedin}, N.Y.},
  \bibinfo{author}{{Hamilton}, A.J.S.}, \bibinfo{year}{2005}.
\newblock \bibinfo{title}{{VOBOZ: an almost-parameter-free halo-finding
  algorithm}}.
\newblock \bibinfo{journal}{\mnras} \bibinfo{volume}{356},
  \bibinfo{pages}{1222--1232}.
\newblock \DOIprefix\doi{10.1111/j.1365-2966.2004.08505.x},
  \href{http://arxiv.org/abs/astro-ph/0402346}{\tt arXiv:astro-ph/0402346}.
\bibitem[{Okabe et~al.(2000)Okabe, Boots, Sugihara and Chiu}]{9780471986355}
\bibinfo{author}{Okabe, A.}, \bibinfo{author}{Boots, B.},
  \bibinfo{author}{Sugihara, K.}, \bibinfo{author}{Chiu, S.N.},
  \bibinfo{year}{2000}.
\newblock \bibinfo{title}{Spatial Tessellations: Concepts and Applications of
  Voronoi Diagrams}.
\newblock \bibinfo{publisher}{Wiley}.
\bibitem[{{Pelupessy} et~al.(2003){Pelupessy}, {Schaap} and {van de
  Weygaert}}]{2003A&A...403..389P}
\bibinfo{author}{{Pelupessy}, F.I.}, \bibinfo{author}{{Schaap}, W.E.},
  \bibinfo{author}{{van de Weygaert}, R.}, \bibinfo{year}{2003}.
\newblock \bibinfo{title}{{Density estimators in particle hydrodynamics. DTFE
  versus regular SPH}}.
\newblock \bibinfo{journal}{\aap} \bibinfo{volume}{403},
  \bibinfo{pages}{389--398}.
\newblock \DOIprefix\doi{10.1051/0004-6361:20030314},
  \href{http://arxiv.org/abs/astro-ph/0303071}{\tt arXiv:astro-ph/0303071}.
\bibitem[{Peterka et~al.(2014)Peterka, Morozov and
  Phillips}]{Peterka:2014:HCD:2683593.2683702}
\bibinfo{author}{Peterka, T.}, \bibinfo{author}{Morozov, D.},
  \bibinfo{author}{Phillips, C.}, \bibinfo{year}{2014}.
\newblock \bibinfo{title}{High-performance computation of distributed-memory
  parallel 3d voronoi and delaunay tessellation}, in:
  \bibinfo{booktitle}{Proceedings of the International Conference for High
  Performance Computing, Networking, Storage and Analysis},
  \bibinfo{publisher}{IEEE Press}, \bibinfo{address}{Piscataway, NJ, USA}. pp.
  \bibinfo{pages}{997--1007}.
\newblock \URLprefix \url{http://dx.doi.org/10.1109/SC.2014.86},
  \DOIprefix\doi{10.1109/SC.2014.86}.
\bibitem[{{Planck Collaboration} et~al.(2014){Planck Collaboration}, {Ade},
  {Aghanim}, {Armitage-Caplan}, {Arnaud}, {Ashdown}, {Atrio-Barandela},
  {Aumont}, {Baccigalupi}, {Banday} and et~al.}]{2014A&A...571A..16P}
\bibinfo{author}{{Planck Collaboration}}, \bibinfo{author}{{Ade}, P.A.R.},
  \bibinfo{author}{{Aghanim}, N.}, \bibinfo{author}{{Armitage-Caplan}, C.},
  \bibinfo{author}{{Arnaud}, M.}, \bibinfo{author}{{Ashdown}, M.},
  \bibinfo{author}{{Atrio-Barandela}, F.}, \bibinfo{author}{{Aumont}, J.},
  \bibinfo{author}{{Baccigalupi}, C.}, \bibinfo{author}{{Banday}, A.J.},
  \bibinfo{author}{et~al.}, \bibinfo{year}{2014}.
\newblock \bibinfo{title}{{Planck 2013 results. XVI. Cosmological parameters}}.
\newblock \bibinfo{journal}{A\&A} \bibinfo{volume}{571}, \bibinfo{pages}{A16}.
\newblock \DOIprefix\doi{10.1051/0004-6361/201321591},
  \href{http://arxiv.org/abs/1303.5076}{\tt arXiv:1303.5076}.
\bibitem[{Rycroft(2009)}]{voro}
\bibinfo{author}{Rycroft, C.H.}, \bibinfo{year}{2009}.
\newblock \bibinfo{title}{Voro++: A three-dimensional voronoi cell library in
  c++}.
\newblock \bibinfo{journal}{Chaos} \bibinfo{volume}{19}.
\newblock \DOIprefix\doi{http://dx.doi.org/10.1063/1.3215722}.
\bibitem[{{Schaap} and {van de Weygaert}(2000)}]{2000A&A...363L..29S}
\bibinfo{author}{{Schaap}, W.E.}, \bibinfo{author}{{van de Weygaert}, R.},
  \bibinfo{year}{2000}.
\newblock \bibinfo{title}{{Continuous fields and discrete samples:
  reconstruction through Delaunay tessellations}}.
\newblock \bibinfo{journal}{\aap} \bibinfo{volume}{363},
  \bibinfo{pages}{L29--L32}.
\newblock \href{http://arxiv.org/abs/astro-ph/0011007}{\tt
  arXiv:astro-ph/0011007}.
\bibitem[{{Springel}(2005)}]{2005MNRAS.364.1105S}
\bibinfo{author}{{Springel}, V.}, \bibinfo{year}{2005}.
\newblock \bibinfo{title}{{The cosmological simulation code GADGET-2}}.
\newblock \bibinfo{journal}{\mnras} \bibinfo{volume}{364},
  \bibinfo{pages}{1105--1134}.
\newblock \DOIprefix\doi{10.1111/j.1365-2966.2005.09655.x},
  \href{http://arxiv.org/abs/astro-ph/0505010}{\tt arXiv:astro-ph/0505010}.
\bibitem[{{Springel}(2010)}]{2010MNRAS.401..791S}
\bibinfo{author}{{Springel}, V.}, \bibinfo{year}{2010}.
\newblock \bibinfo{title}{{E pur si muove: Galilean-invariant cosmological
  hydrodynamical simulations on a moving mesh}}.
\newblock \bibinfo{journal}{\mnras} \bibinfo{volume}{401},
  \bibinfo{pages}{791--851}.
\newblock \DOIprefix\doi{10.1111/j.1365-2966.2009.15715.x},
  \href{http://arxiv.org/abs/0901.4107}{\tt arXiv:0901.4107}.
\bibitem[{{Starinshak} et~al.(2014){Starinshak}, {Owen} and
  {Johnson}}]{2014CoPhC.185.3204S}
\bibinfo{author}{{Starinshak}, D.P.}, \bibinfo{author}{{Owen}, J.M.},
  \bibinfo{author}{{Johnson}, J.N.}, \bibinfo{year}{2014}.
\newblock \bibinfo{title}{{A new parallel algorithm for constructing Voronoi
  tessellations from distributed input data}}.
\newblock \bibinfo{journal}{Computer Physics Communications}
  \bibinfo{volume}{185}, \bibinfo{pages}{3204--3214}.
\newblock \DOIprefix\doi{10.1016/j.cpc.2014.08.020}.
\bibitem[{{Voronoi}(1908)}]{voronoi}
\bibinfo{author}{{Voronoi}, G.}, \bibinfo{year}{1908}.
\newblock \bibinfo{journal}{Z. Reine Angew. Math} \bibinfo{volume}{134},
  \bibinfo{pages}{198}.
\bibitem[{Wu et~al.(2011)Wu, Gonzalez, Lan, Gnedin, Kravtsov, Rudd and
  Yu}]{Wu2011}
\bibinfo{author}{Wu, J.}, \bibinfo{author}{Gonzalez, R.E.},
  \bibinfo{author}{Lan, Z.}, \bibinfo{author}{Gnedin, N.Y.},
  \bibinfo{author}{Kravtsov, A.V.}, \bibinfo{author}{Rudd, D.H.},
  \bibinfo{author}{Yu, Y.}, \bibinfo{year}{2011}.
\newblock \bibinfo{title}{Performance emulation of cell-based {AMR} cosmology
  simulations}, in: \bibinfo{booktitle}{2011 {IEEE} International Conference on
  Cluster Computing}, \bibinfo{publisher}{Institute of Electrical {\&}
  Electronics Engineers ({IEEE})}.
\newblock \URLprefix \url{http://dx.doi.org/10.1109/CLUSTER.2011.10},
  \DOIprefix\doi{10.1109/cluster.2011.10}.
\bibitem[{{York} et~al.(2000){York}, {Adelman}, {Anderson}, {Anderson},
  {Annis}, {Bahcall}, {Bakken} and {SDSS Collaboration}}]{2000AJ....120.1579Y}
\bibinfo{author}{{York}, D.G.}, \bibinfo{author}{{Adelman}, J.},
  \bibinfo{author}{{Anderson}, Jr., J.E.}, \bibinfo{author}{{Anderson}, S.F.},
  \bibinfo{author}{{Annis}, J.}, \bibinfo{author}{{Bahcall}, N.A.},
  \bibinfo{author}{{Bakken}, J.A.}, \bibinfo{author}{{SDSS Collaboration}},
  \bibinfo{year}{2000}.
\newblock \bibinfo{title}{{The Sloan Digital Sky Survey: Technical Summary}}.
\newblock \bibinfo{journal}{\aj} \bibinfo{volume}{120},
  \bibinfo{pages}{1579--1587}.
\newblock \DOIprefix\doi{10.1086/301513},
  \href{http://arxiv.org/abs/astro-ph/0006396}{\tt arXiv:astro-ph/0006396}.

\end{thebibliography}


\end{document}